# Trade Facilitation and Economic Growth Among Middle-Income Countries

**Victor Ushahemba Ijirshar**
Department of Economics,
Benue State University, Makurdi-Nigeria

**Abstract**
This study examined the relationship between trade facilitation and economic growth among the middle-income countries from 2010 to 2020 using 94 countries made up of 48 lower-middle-income countries and 46 upper-middle-income countries. The study utilized both difference and system Generalised Method of Moments (GMM) since the cross-sections (N) were greater than the periods (T). The study found that container port traffic, quality of trade and transport-related infrastructure have a strong influence on imports and exports of goods and national income while trade tariff hurts the growth of the countries. The study also found that most of the trade facilitation indicators indicated a weak positive influence on trade flows and economic growth. Based on these findings, the study recommends that reforms aimed at significantly lowering the costs of trading across borders among middle-income countries should be highly prioritized in policy formulations, with a focus on the export side by reducing at-the-border documentation, time, and real costs of trading across borders while the international organizations should continue to report the set of Trade Facilitation Indicators (TFIs) that identify areas for action and enable the potential impact of reforms to be assessed.

**Keywords:** Economic Growth, Goods Exports, Goods Imports, Trade Facilitation

## 1.1 Introduction

As World trade becomes more liberalized including lower tariffs and quotas, the focus of policymakers has shifted to the impediments to the cross-border movement of goods, particularly to those of an administrative and logistical nature. The transport connection, quality of logistical services, and border management all play growing roles in the determination of international trade flows and economic growth. Trade facilitation in particular has been identified as a tool for increased and smoother cross-border trade (United Nations, 2016).

In simple terms, trade facilitation is the easing of the flow of goods across borders. It simplifies and harmonizes international trade procedures including the activities, practices and formalities involved in collecting, presenting, communicating, and processing data required for the movement of goods at the international level. It is the reduction of transaction costs associated with institutional trade barriers. The primary goal of trade facilitation is to help make trade across borders (imports and exports) faster, cheaper, and more predictable while ensuring its safety and security. Thus, it helps streamline policies and procedures required in conveying goods from one country to another to achieve the stated goal. The argument is that trade facilitation helps to enhance trade flows (Florensa, Márquez-Ramos & Recalde, 2015; Ma´rquez-Ramos & Martinez-Gomez, 2014; Amoako-Tuffour, Balchin, Calabrese & Mendez-Parra, 2016), it is expected also to stimulate economic growth of the middle-income countries. To United Nations (2021), trade facilitation reduces unnecessary delays, attracts investments, enhance the speed of operations and lower transaction costs, supports job creation, and economic growth. Based on this argument, the attention of scholars, policymakers, and the community involved in international trade has recently focused on investigating the extent to which trade facilitation can influence trade flows to have its largest positive effect on economic growth.

The middle-income countries are the emerging and developing economies. They are defined as lower-middle-income economies (those with a GNI per capita between $1,036 and $4,045); and upper-middle-income economies (those with a GNI per capita between $4,046 and $12,535). They form 75% of the world's population, 62% of the world's poor, one-third of global GDP and are major engines of global growth (World Bank, 2021). For many middle-income countries, an increased level of integration into the global economy is a key driver of productivity and growth (Safaeimanesh & Jenkins, 2021; Ijirshar, 2019). Recent studies from the World Trade Organization (WTO) suggest that improvements in border administration throughout the world could boost global trade by US$ 1 trillion per year, meaning that trade facilitation could have a bigger impact on international trade than if all the world's remaining tariffs were removed (International Chamber of Commerce, 2021). Larger gains are expected from developing or middle-income countries since they have greater scope for improvement to the trade environment and the existence of some form of diminishing returns in investments in infrastructure that will facilitate trade. Granted that middle-income countries lag behind the developed economies as far as the quality of the trade environment is concerned, one could hypothesize that countries under such category can reap greater benefits from trade facilitation reforms. Trade facilitation helps the developing countries or middle-income countries where it frequently takes three times as many days to export goods as it does in developed ones, requiring nearly twice as many documents, and six times as many signatures (United Nations, 2021; World Bank, 2018). In a bid to enhance the general benefits, reduce trade costs, and boost global trade, the trade facilitation agreement was initiated and implemented to reap the full benefits of trade facilitation on 22 February 2017 after two-thirds of the World Trade Organisation (WTO) membership completed their domestic ratification process (WTO, 2021).

Trade facilitation has three levels of operations: at the national, regional and international levels. While at the regional and international levels, standards and agreements are developed and agreed upon, the operational implementation of trade facilitation measures including these standards, takes place at the national level. Hence, its operations at the national level have effects on trade flows of an economy and economic growth especially developing economies as noted earlier. Besides, trade allows countries to specialize; enables technological know-how and ideas to spread; promotes competition; and yields economies of scale that in turn foster economic growth (Anjande, Ijirshar, Asom, Akiri & Sokpo, 2020; Ijirshar, 2019).

Theoretically, the benefits of economic growth accrued from trade are made explicit by both extended standard neoclassical exogenous and endogenous economic growth models (Sakyi, Villaverde, Maza & Bonuedi, 2017). These theories, which link trade to growth through endogenous technological change and technological diffusion, posit that using transfer of knowledge and technology, trade boosts economic growth (Sakyi, Villaverde, Maza & Bonuedi, 2017; Kummer-Noormamode, 2014; Sakyi, Villaverde & Maza, 2015; Brueckner & Lederman, 2015).

Statistics have shown that exports of goods and services (current US$) by the middle-income countries increased from 1.502 trillion in 2000 to 6.874 trillion in 2020 with little fluctuations while the imports of goods and services (current US$) by the middle-income countries increased from 1.401 trillion in 2000 to 6.601 trillion in 2020. However, the export of goods and services as percent GDP was 26.19% in 2000 to 22.453% in 2020 with a peak of 32.929% in 2006. In terms of gross national income, the middle-income countries recorded about 17.826 trillion in 2000 and 68.789 trillion in 2020 with an improved pattern over time. However, the logistic performance indices show that there is an unstable level of efficiency of the customs clearance process, poor quality of trade and transport-related infrastructure, lack of competence and quality of logistics services, ease of arranging competitively priced shipments, among others (World Bank, 2021).

In the previous empirical studies, the effects of trade facilitation on economic growth have been broadly studied in the literature for both developed and developing countries. Notably, most authors argued that through the reduction of tariffs and institutional trade barriers, countries are open more to trade and tend to perform remarkably better than less open economies (Safaeimanesh & Jenkins, 2021; Sakyi, Villaverde, Maza & Bonuedi, 2017). Conversely, the burdensome customs requirements present real challenges to firms of all sizes to trade internationally particularly for Small and Medium-Sized Enterprises (SMEs) in developing or middle-income countries. However, the examination of the influence of trade facilitation on trade flows (while disposing the effects on exports and imports) and consequently national income of countries especially the middle-income countries that are expected to take much advantage from the trade facilitation has not been investigated. It is in view of this that this study intends to examine the trade facilitation efforts over time and how these trade facilitation measures have influenced the level of trade flows and national income of the middle-income countries. The study covers 94 middle-income countries from 2010 to 2020. This period coincides with the aftermath of the global financial crisis of 2007-2008.

The rest of this paper is organised as follows; section 2 discusses review of related literature. The methodology is presented in section 3 while section 4 presents, discusses and interprets the empirical results. Section 5 offers conclusion and policy recommendations.

## 2. Review of Related Literature

### 2.1 *Theoretical Review*

Generally, international trade theories can be classified broadly into descriptive theories and prescriptive theories. Descriptive theories are theories that explain what pattern of trade that exists, why trade occurs and what kinds of products are to be traded from country to country. But prescriptive theories are international trade theories that explain how issues relating to international trade should be done, to what extent should government control or influence trade through cross border policies, whether there should be a limit on the amount or the type of goods that are traded and the countries that trade should occur. First, the Mercantilism theory originated the idea of trade around the 16th to 18th century. They explained trade in terms of wealth; that a country's wealth is measured by its holdings of treasure. This is one of the prescriptive theories. It prescribes that countries should export more than they import to amass treasure (Mun, 1664). That is a favourable balance of trade or trade surplus. The theory also suggested that countries should import raw materials and export higher valued products.

But arguing on the notion of the country's wealth, Adam smith towards the end of 18 century (1776), described the theory as being misguided and the concept of the country's wealth as an illusion (Frederick, 2018). Adam Smith posits that country's wealth is based on its available goods and services rather than on gold (Adam Smith Institute, 2021). The theory argued that trade is beneficial through specialization since it increases efficiency, hence, it should be unrestricted. The theory asserts that countries should produce the goods that they have an advantage in producing. The advantages include natural advantages such as natural resources, climate, geographical location (transportation cost); acquired advantages such as skills and technology. However, Ricardo questioned the conclusion by Adam Smith that there is no mutually beneficial trade if a country is at a disadvantage in the production of both commodities. Ricardo stated that there are gains from trade even when a country has no absolute advantages with its trading partners. The theory states that a country should specialize in the goods or services it can produce at the lowest opportunity cost and trade with another country (Thompson, 2017). The theory also argues that for the trade to be beneficial, the exchange rate must lie between the ratios of the countries' opportunity costs. The contribution of trade to growth varies depending on whether the force of comparative advantage directs the economy's resources towards activities that generate long-run growth or away from such activities. Moreover, theories suggest that, due to technological or financial constraints, developing or middle-income countries may lack the social capability required to adopt technologies developed in more advanced economies (Ijirshar, 2019). Thus, the growth effect of trade may differ according to the level of trade facilitation.

But the quantity and quality of the factor endowment influence the commodity countries have a comparative advantage (Factor endowment approach) (Heckscher-Ohlin). However, the Heckscher-Ohlin theory assumed that technology is very rigid (constant) (Fincke, 2006).

The new trade theory also argued that trade can take place between similar countries. Krugman in the New Trade Theory (NTT) came with a new version of the transactions at the international trade (Wangwe, 1993). The theory is a collection of economic models in international trade which focuses on the role of increasing returns to scale and the network effects. The theory relaxed the assumption of constant returns to scale and regarded the comparative advantage theory as old trade theory. The network effect is the effect that one user of a good or service has on the value of the product to others without the intention to create the value for others (Bandwagon effect) (Hill & Arun, 2014). Even though countries may be completely the same, they may have an incentive to trade. This can be attributed to the trade facilitation components.

In terms of economic growth, the study employed Solow–Swan neo-classical growth model that was introduced by Solow (1956) and Swan (1956). The essence of the Solow–Swan neoclassical growth theory is that the potential rate of growth of output which represents the equilibrium and 'natural' rates of growth -is determined exogenously by the rate of growth of the labour force and technological progress. According to Solow (1956), Output along the balanced growth path is determined by technology, investment rate and the population growth rate and that growth in output and the volume of international trade are closely related. Thus, changes in trade-related issues orchestrated by technology may affect the level of national income.

The consideration of the effects of trade facilitation in trade theory began with the development of the 'iceberg' method (Irarrazabal, Moxnes & Opromolla, 2010; Samuelson, 1954). Samuelson (1954) utilized this theory in explaining the influence of transportation costs, or the possible effects of transport impediments on trade. Several studies have used the iceberg theory to analyse the impacts of trade costs that arise due to insufficient trade procedures (poor trade facilitation), using partial equilibrium models as well as general equilibrium models (Safaeimanesh & Jenkins, 2021; Perera, Siriwardana & Mounter, 2017). The theoretical developments of the effects of trade facilitation in a general equilibrium framework can be discussed in terms of both classical trade and new trade theory. The classical trade theory consists of the Ricardian model and the Heckscher-Ohlin theory. These two theories explain that countries produce goods having a comparative advantage due to relative productivity differences (comparative technological advances) or endowments of factors of production (use abundant factors of production more intensively), respectively.

In the Ricardian theory, if countries do not trade with each other (in autarky), the relative price of one good expressed in terms of the other good differs between them. This motivates the enhancement of bilateral trade, as the world market price exceeds the domestic prices due to the specialised production of the good that has a comparative advantage. However, the insufficient trade facilitation lowers the difference between the relative prices faced by both trading partners and the international price moves closer to the autarky price, leading to reduced trade and consumption, as well as economic growth and vice versa, ceteris paribus. On the other hand, assuming similar productivity in both countries, the Heckscher-Ohlin model describes the differences in factor endowments. The model assumes that there are two factors in production, labour and capital. One country is labour abundant and the other capital abundant. The prices of two goods in two countries differ in autarky due to different factor endowments. The labour-abundant country produces a labour-intensive good with a domestic price lower than the foreign price.

According to the New Trade Theory, trade costs can have a disproportionately negative impact on developing countries. With constant returns to scale and a small manufacturing sector, developing countries produce more agricultural or natural resource-related goods. In contrast, developed countries have a large manufacturing sector that operates on the principle of increasing returns to scale. Trade costs can reduce trade in both developed and developing countries, resulting in a disproportionate reallocation of manufacturing goods to developed countries and agricultural and natural resources to developing nations. This emphasizes the importance of lowering trade costs in both developed and developing countries in order to diversify trade.

## 2.2 *Empirical Studies on Trade Facilitation and Economic Growth*

Assessing the potential annual economic gain to be had from trade facilitation by the coastal countries of the Economic Community of West African States (ECOWAS), Safaeimanesh and Jenkins (2021) used a partial equilibrium welfare economics framework by employing a set of export supply and import demand elasticities for each country that are derived using a general equilibrium estimation method. The study found that the economic welfare benefits resulting from a reduction in excessive import and export border and documentary compliance costs are considerable for ECOWAS countries. The study particularly focused on the economic gain from export supply of countries when there is trade facilitation.

Sakyi, Villaverde, Maza, and Bonuedi (2017) also studied the effects of trade and trade facilitation on African economic growth. The study assessed trade facilitation using three indicators: trade, export, and import-related costs derived from principal component analysis. To address potential endogeneity concerns, the study estimated an augmented growth model using a dynamic system GMM estimation technique. According to Sakyi, Villaverde, Maza, and Bonuedi (2017), trade facilitation is an important channel through which trade influences economic growth. Incorporating the effects of trade facilitation on both trade and economic growth, Perera, Siriwardana and Mounter (2017) used South Asian annual data. The study revealed that poor trade facilitation restricts trade between countries as it increases Trade Transaction Costs (TTCs). Trade delays are relatively high and affect the region's landlocked countries even more adversely in the region. An efficiently facilitated trading system enables these countries to participate more actively in global trade.

## 2.3 *Empirical Studies on Trade Facilitation and Trade Flows*

Assessing the effects of trade facilitation on trade flows, Ali and Shakoor (2020) examined the relationship between trade facilitation indicators and trade flow in Asian countries. The study used the main trade facilitation variables such as tariffs, time to import/export and cost to import/export, population, and ICT from secondary sources in 44 Asian countries. The study used fixed-effect fixed and random effect methodology for estimation and then used the Husman test to choose between fixed and random effect results. Findings showed that there is a significant impact of trade facilitation on imports and exports volume and it clarifies that tariff and number of documents to import and export hurts trade flow.

Yu and Luu (2020) also examined the impact of trade facilitation on trade flows in Vietnam from 2007 to 2018 in two scenarios: with and without Free Trade Agreements (FTAs). The gravity model and the Generalized Method of Moments were used to analyze Vietnam's import and export data sets from that time period (GMM). According to Yu and Luu (2020), trade facilitation had a positive impact on Vietnam's trade flows, and its effect was consistent with and without FTAs. In related finding and further assessing the role of free trade agreements, Seetanah, Sannassee, and Fauzel (2016) investigated the impact of trade facilitation on trade flows in a sample of 20 African economies from 2007 to 2014. The study discovered, using a panel vector autoregressive framework, that trade facilitation improves trade flows in those African countries.

In decomposing the effects of trade facilitation indicators on agricultural and non-agricultural products, Mai and Ngoc (2019) used the structural gravity model with the Poisson Pseudo maximum likelihood estimation approach to assess the correlation between these indicators and trade value. The dataset records the trade flow between Vietnam and 22 strategic partners on a product-by-product basis. According to the study's findings, trade facilitation and logistics infrastructure have had a direct impact on trade flows across partners and vary for agricultural and non-agricultural products.

In the same vein, Sá Porto, Canuto, and Morini (2015) examined the effects of selected trade facilitation measures on international trade flows for 72 countries from 2011 to 2012. Four equations were estimated using a gravity model: a pooled cross-section model, a fixed-effects model, a random effects model, and a Poisson maximum likelihood estimator. The research found that having an authorized economic operator program and having a single-window program improved countries' trade performance, and that trade facilitation measures in general helped countries improve their trade performance.

Odularu and Alege (2019) used the gravity model to investigate the impact of administrative barriers on bilateral trade flows in member countries of the Economic Community of West African States (ECOWAS). For the study, a dynamic gravity model was estimated, and the results were mixed. The study also discovered that key independent variables had varying degrees of statistical significance. The Poisson Pseudo-Maximum Likelihood (PPML) estimates, on the other hand, revealed that the independent variables are not statistically significant.

In trying to assess the individual effects of trade facilitation on trade flows, Wilson, Mann, and Otsuki (2003) constructed indicators for measuring trade facilitation using country-specific data for port efficiency, customs environment, regulatory environment, and e-business usage in the Asia Pacific region. A gravity model with tariffs and other standard variables was used to estimate the relationship between these indicators and trade flows. According to the study, improved port efficiency has a large and positive effect on trade flows, and improvements in customs and increased e-business use significantly expand trade, but to a lesser extent than improvements in ports or regulations. Regulatory barriers, on the other hand, stymie trade. It was also revealed that the benefits of specific trade facilitation efforts are estimated by quantifying discrepancies in these four areas among Asia Pacific Economic Cooperation members (APEC).

### 2.4 *Empirical Studies on Trade Facilitation and the General Economy*

Examining the influence of trade facilitation on the general economy, Avetisyan and Hertel (2021) analyzed the effect of improved global logistics and trade facilitation on transport mode choice in international trade using the Global Trade Analysis Project (GTAP) model of global trade incorporating modal choice. The study found that the quality of logistics infrastructure influences modal choice in international trade with improved logistics performance generally leading to increased reliance on air transportation. The study also found that improvement in logistics performance index in the poorest countries of the world reduces the overall cost of transport and amount of services required to transport a given product along a given route by a given mode. Kouty (2021) also examined whether the implementation of AfCFTA cannot be done without harmonized trade procedures. Using a gravity model of 49 African countries over the 2010-2015, the study estimated the impact of inefficient trade procedures on intra-African trade and found that trade procedures such as the number of documents required to import goods and border compliance negatively affect intra-African trade.

Seck (2016) examined how firms, relative to their counterparts in the developing world, would respond to changes in the trade environment as a result of trade facilitation reforms. Using data from World Bank's Enterprise Surveys, the study suggests that improving customs clearance, government regulations, trade finance, and energy and telecommunication infrastructure contributes to increasing the probability of firms' entry into exporting and importing, as well as to the extent of their trade. The study further showed that African firms tend to respond more to a changing environment, owing to the greater constraints that they face while exports tend to be more responsive than imports, suggesting a favourable short-term adjustment of the balance of payments.

# 3 Methodology

## 3.1 Data Description

The panel data utilised in this study include data on efficiency of customs and border management, the burden of customs procedure, port container traffic, the logistics performance indices such as ability to track and trace consignments, competence and quality of logistics services, ease of arranging competitively priced shipments, efficiency of customs clearance process, frequency with which shipments reach consignee within scheduled or expected time, and the quality of trade and transport-related infrastructure, the overall logistics performance index, the quality of port infrastructure, tariff rate (applied, weighted mean) for all products, goods exports, goods imports, foreign direct investment, gross fixed capital formation, the gross national income. Data on all the logistics performance indices are ranked from 1 (low) to 5 (high). Data on the efficiency of customs and border management is ranked data from extremely underdeveloped=1 to well developed and efficient by international standards=7. The data on the burden of customs procedure is also ranked data from extremely inefficient=1 to extremely efficient=7 likewise the data on quality of port infrastructure. The data on the efficiency of customs and border management and the burden of customs procedures are sourced from World Economic Forum Global Competitiveness Report. Airfreight transportation measures the volume of freight, express, and diplomatic bags carried on each flight stage measured in metric tons times kilometres travelled. Port container traffic measures the flow of containers from land to sea transport modes, and vice versa, in twenty-foot equivalent units. The data on air freight transportation and port container traffic is sourced from International Civil Aviation Organization and Containerisation International Yearbook respectively. Data on average time to clear exports through customs (days), the efficiency of the customs clearance process (Logistics performance index), tariff rate (applied, weighted mean) for all products, goods exports, goods imports, foreign direct investment, gross fixed capital formation, and Gross Nation Income (PPP) are sourced from World Bank. The data covers the period 2010 to 2020. The study used panel data on 94 middle-income countries comprising of 48 lower-middle-income countries and 46 upper-middle-income countries. The lower-middle-income countries are: Algeria, Angola, Bangladesh, Benin, Bhutan, Bolivia, Cabo Verde, Cambodia, Cameroon, Comoros, Congo Republic, Cote d'ivoire, Djibouti, Egypt Arab Republic, El Salvador, Ghana, Haiti, Honduras, India, Indonesia, Iran Islamic Republic, Kenya, Kyrgyz Republic, Lao PDR, Lesotho, Mauritania, Mongolia, Morocco, Myanmar, Nepal, Nicaragua, Nigeria, Pakistan, Papua New Guinea, Philippines, Sao Tome and Principe, Senegal, Solomon Islands, Sri Lanka, Tajikistan, Tanzania, Timor-Leste, Tunisia, Ukraine,

Uzbekistan, Vietnam, Zambia, and Zimbabwe. The upper-middle-income countries are: Albania, Argentina, Armenia, Azerbaijan, Belarus, Bosnia and Herzegovina, Botswana, Brazil, Bulgaria, China, Colombia, Costa Rica, Cuba, Dominican Republic, Ecuador, Equatorial Guinea, Fiji, Gabon, Georgia, Guatemala, Guyana, Iraq, Jamaica, Jordan, Kazakhstan, Lebanon, Libya, Malaysia, Maldives, Mauritius, Mexico, Moldova, Montenegro, Namibia, North Macedonia, Panama, Paraguay, Peru, Romania, Russian Federation, Serbia, South Africa, Suriname, Thailand, Turkey, and Turkmenistan. These data were obtained from The World Bank.

### 3.2 Model Specification

This study adopts the Ricardian theory of comparative advantage for modelling the effect of trade on national income. The study also incorporates four determinants of national income in the Solow-Swan theory. According to the Ricardian theory of comparative advantage, trade is mutually beneficial even if one country has a comparative disadvantage in the production of both commodities of trade (Rehim, 2002). This can be captured in a functional form as:

$$GNI_{it} = f(TF_{it}) \quad - \; - \; - \; - \; - \; - \; - \; - \; - \quad (1)$$

Where GNI= Gross National Income, and TF=Trade Facilitation

Given that economic growth or gross national income is also determined by basic variables such as foreign direct investment, gross fixed capital formation, exports and imports of goods (Anjande, Ijirshar, Asom, Akiri & Sokpo, 2020; Ijirshar, 2019), the model can be stated in a definitional form with semi-transformation as:

$$\ln GNI_{it} = f\left(ATCE_{it}, \ln PCT_{it}, AFT_{it}, QPI_{it}, LPI_{it}, TRF_{it}, \ln EXPG_{it}, \ln IMPG_{it}, \ln FDI_{it}, \ln GFCF_{it}\right) \quad (2)$$

In determining the effects of trade facilitation on trade flows in Africa, the study used the iceberg theory. It explains how poor trade facilitation affect trade flows (Safaeimanesh & Jenkins, 2021; Perera, Siriwardana & Mounter, 2017). The model can be stated as:

$$Trade_{it} = f(TF_{it}) \quad (3)$$

Decomposing trade flows into exports and imports of goods (since trade facilitation affects mostly the movement of goods across borders) and further incorporating other determinants of trade flows, the equation can be restated with semi-transformation as:

$$\ln EXPG_{it} = f\left(ATCE_{it}, \ln PCT_{it}, AFT_{it}, QPI_{it}, LPI_{it}, TRF_{it}, \ln FDI_{it}, \ln GFCF_{it}\right) \quad (4)$$

$$\ln IMPG_{it} = f\left(ATCE_{it}, \ln PCT_{it}, AFT_{it}, QPI_{it}, LPI_{it}, TRF_{it}, \ln FDI_{it}, \ln GFCF_{it}\right) \quad (5)$$

Following a typical dynamic (including a lag of the dependent variable as a regressor in a model makes it a dynamic model) panel data model, a panel Generalized Method of Moment (panel GMM) is specified as:

$$Y_{it} = \delta Y_{i,t-1} + X'_{it}\beta + u_i + \eta_{it} \quad (6)$$

And the generalized framework as:

$$Y_{it} = \delta w_{it} + X'_{it}\beta + \varepsilon_{it} \quad (7)$$

$$i = 1, \ldots, n; \quad t = 1, \ldots, T$$

$\varepsilon_{it} = u_i + \eta_{it}$

Where

$w_{it}$ is a vector of predetermined covariates (which may include the lag of Y) and endogenous covariates, all of which were correlated with the $u_i$ (error term that captures all other omitted factors) and note; $\delta$ measures the causal effect of lagged dependent variable on current levels of the dependent variable.

Equation (4) is therefore re-stated as:

$$\ln EXPG_{it} = \delta EXPG_{i,t-1} + \beta_1 ATCE_{it} + \beta_2 \ln PCT_{it} + \beta_3 AFT_{it} + \beta_4 QPI_{it} + \beta_5 LPI_{it} + \beta_6 TRF_{it} + \beta_7 \ln FDI_{it} + \beta_8 \ln GFCF_{it} + \varepsilon_{it} \quad (8)$$

Where, $\beta_1 - \beta_8 =$ Parameter Coefficients to be estimated, $\varepsilon_{it} =$ Mutually Independent idiosyncratic error.

Equation (5) is therefore re-stated as:

$$\ln IMPG_{it} = \delta IMPG_{i,t-1} + \beta_1 ATCE_{it} + \beta_2 \ln PCT_{it} + \beta_3 AFT_{it} + \beta_4 QPI_{it} + \beta_5 LPI_{it} + \beta_6 TRF_{it} + \beta_7 \ln FDI_{it} + \beta_8 \ln GFCF_{it} + \varepsilon_{it} \quad (9)$$

Where, $\beta_1 - \beta_8 =$ Parameter Coefficients to be estimated, $\varepsilon_{it} =$ Mutually Independent idiosyncratic error.

Equation (2) is therefore re-stated as:

$$\ln GNI_{it} = \delta \ln GNI_{i,t-1} + \beta_1 ATCE_{it} + \beta_2 \ln PCT_{it} + \beta_3 AFT_{it} + \beta_4 QPI_{it} + \beta_5 LPI_{it} + \beta_6 TRF_{it} + \beta_7 \ln EXPG_{it} + \beta_8 \ln IMPG_{it} + \beta_9 \ln FDI_{it} + \beta_{10} \ln GFCF_{it} + \varepsilon_{it} \quad (10)$$

Where, $\beta_1 - \beta_{10} =$ Parameter Coefficients to be estimated, $\varepsilon_{it} =$ Mutually Independent idiosyncratic error

## 3.3 Variables Description and Measurements

A brief description of the variables with their respective measurements and data sources are presented in Table 1.

## 3.4 Estimation Technique

The method of data analysis is the Generalised Method of Moments (GMM). Given that the number of countries (cross-sections) is greater than the number of periods (time dimension), the GMM estimators are efficient. The period of 2010–2020 is considered adequate to examine variables' dynamic effects on the relationship under study using both the difference and system Generalised Method of Moments (GMM) estimators within a dynamic panel framework. The consideration of this approach is informed by two basic sources of persistence over time; first, autocorrelation resulting from the inclusion of a lagged variable among the explanatory variables and second, the unobserved main effects and interaction effect characterising the heterogeneity among the units (Olubusoye, Salisu & Olofin, 2016; Baltagi, 2008). Hence, applying either OLS (Ordinary Least Squares) or Fixed Effects (FE) estimator may render the estimates biased and inconsistent.

The Difference GMM is the estimation that proceeds after first-differencing the data in order to eliminate the fixed effects while the system GMM augments Difference GMM by estimating simultaneously in differences and levels. The Arellano–Bond (Arellano & Bond, 1991) and Arellano–Bover/Blundell–Bond (Arellano & Bover, 1995; Blundell & Bond, 1998) dynamic panel estimators are increasingly popular. Both are general estimators designed for situations with "small T and large N" panels, meaning few time periods and many countries as obtainable in this study; a linear functional relationship; one left-hand-side variable that is dynamic, depending on its own past realizations; independent variables that are not strictly exogenous, meaning they are correlated with past and possibly current realizations of the error; fixed individual effects; and heteroskedasticity and autocorrelation within individuals but not across them. The cross sectional dependence was used in testing errors across the panels.

The dynamic panel model known as Blundell and Bond (1998) system GMM estimator was therefore also employed. This is an improvement of the standard (Arellano and Bond) GMM estimators. The system GMM estimator by Blundell and Bond (1998) showed that an additional mild stationary restriction on the initial conditions process allows the use of an extended (system) GMM estimator that uses lagged differences of $GNI_{it}$, $EXPG_{it}$, and $IMPG_{it}$, as instruments for equations at levels, in addition to lagged levels of $GNI_{it}$, $EXPG_{it}$, and $IMPG_{it}$ as instruments for equations in first differences (Balatgi, 2008; Olubusoye, Salisu & Olofin, 2016).

The Blundell and Bond system GMM estimator adopted, the levels or untransformed equation is still instrumented with differences and describe how the original equation in levels is added to the system (that is, in addition to lagged levels of variables as instruments for equations in first differences. In the system GMM, variables in levels are instrumented with suitable lags of their first differences with the assumption that these differences are uncorrelated with the unobserved country effects. Given that the lags of the dependent variable (and any other variables used as instruments that are not strictly exogenous) are endogenous, thus bad instruments, the study conducted autocorrelation AR(1) and AR(2) tests. This is because based on the theoretical construct, the study expects the presence of the first-order autocorrelation particularly for the dynamic panel data model in first differences while the presence of a higher-order autocorrelation may render some lags invalid as instruments. Sargan statistic and Hasen J Statistic were used to test for the validity of instruments in the system GMM regression estimated. Hence, the difference - in- Sargan/Hansen statistics which test whether subsets of instruments are valid were computed. Both the difference GMM and the system GMM estimators have one-and two-step variants. Both the one-and two-steps were conducted for robustness check.

The study also used Levin-Lin-Chu (2002) unit-root test and Im-Pesaran-Shin (2003) in testing for unit root in the series while Dumitrescu & Hurlin Granger (2012) non-causality test was used in testing for panel causality.

## 4. Results and Discussion

### 4.1 *Descriptive Statistics*

From the descriptive statistics in Table 2, most of the variables such as gross national income, goods exports, goods imports, the efficiency of customs and border management, the burden of customs procedure (captured), Airfreight transportation, port container traffic, the efficiency of the customs clearance process (Logistics performance index), tariff rate (applied, weighted mean) for all products, goods exports, goods imports, foreign direct investment, gross fixed capital formation, the growth rate of real Gross Domestic Product recorded higher output for upper-middle-income countries to lower-middle-income countries except for Average time to clear exports through customs (days) and average tariff rate. This shows that upper-middle-income countries are characterized by healthy indicators than lower-middle-income countries.

### 4.2 *Panel Unit Root Tests Results*

Since unit root tests are necessary for the application of the GMM either with difference moments conditions or system moments conditions, Levin-Lin-Chu (2002) unit-root test and Im-Pesaran-Shin (2003) unit-root tests were used. The results indicate that most panels contain unit roots at levels but after first differencing, they all became stationary (that is all the variables with unit roots were integrated at first difference). The results of Cross Section Dependence (Pesaran CD) shows that errors or disturbances in the panel data regression are cross-sectional independent and hence, the panel data regression is free from the cross-sectional dependence problem.

### 4.3 *Result of Granger non-causality test between Trade Facilitation and Economic Growth*

The results of the Dumitrescu & Hurlin Granger non-causality test are presented in Table 3. The results show a bidirectional relationship between goods exports and gross national income at 5% level of significance likewise that of goods imports and gross national income. The study also found a unidirectional relationship running from gross national income to average time to clear exports through customs, air freight transportation, quality of port infrastructure, competence and quality of logistics services, ability to track and trace consignments, ease of arranging competitively priced shipments, the frequency with which shipments reach consignee within scheduled or expected time, the quality of trade and transport-related infrastructure, logistics performance index and tariff rate at 5% level of significance. There also exists a bidirectional relationship between goods exports and goods imports at least among one of the middle-income countries. From the results in Table 3, there are several causal relationships at least in one of the middle-income countries across the variables incorporated in the models. This explains the endogenous nature of the variables and the need to treat them in a system as applied.

### 4.4 *Impact of Trade Facilitation on Export Goods among Middle-Income Countries*

The result from the one-step difference estimates of the model with the effects of logistic performance indices and the model with the effect of overall logistic performance index shows that container port traffic exert a strong positive influence on exports of goods among the middle-income countries at 5% level of significance, and the strong positive influence of competence and quality of logistics services, and the quality of trade and transport-related infrastructure on the goods exports from the two-step system GMM at 5% level of significance as revealed by the two-step system GMM and the one-step difference GMM as presented in Table 4.

The study also found that the overall logistic performance of customs and border management has a positive influence on the exports of goods of the middle-income countries at 10% level of significance from the difference GMM results of the model with the effect of the overall logistic performance index. The implication is that the trade facilitation measures enhance the exports of goods from the middle-income countries but they may require improvement to exert a stronger influence on the performance of the middle-income countries in terms of exports of goods.

The study also found that the lagged dependent variables are positive and statistically significant at influencing the exports of goods among the middle-income countries. The results show that number of instruments are at most (33) that are less than the number of groups (94). The Sargan and Hansen tests of over-identifying restrictions are valid as expected in supporting the choice of instruments, the AR (2) value of the models estimated are greater than the critical value (0.05) implying that the original error term is serially uncorrelated and the moment conditions are correctly specified. The study, using the Wald chi2 statistic, also found that there is a weak joint influence of all trade facilitation indicators on exports of goods.

### 4.5 *Impact of Trade Facilitation on Goods Imports among Middle-Income Countries*

The results from the one-step difference GMM estimates of the model with the effects of logistic performance indices and the model with the effect of overall logistic performance index show that container port traffic has a significant positive influence on imports of goods among the middle-income countries at 5% level of significance, and the strong positive influence of quality of trade and transport-related infrastructure on the imports of goods as revealed by the one-step and two-step GMM estimates at 5% level of significance as presented in Table 5. The study further revealed that the average time to clear goods through customs discourages imports of goods significantly at 10% level of significance. The study also reveals that tariff rate and the efficiency of the customs clearance process exert a negative influence on the import of goods to middle-income countries as revealed by the one-step GMM estimates in Table 5.

The study also found that the overall logistic performance of customs and border management has a positive influence on the exports of goods of the middle-income countries at 10% level of significance from the difference GMM results of the model with the effect of overall logistic performance index. The implication is that the trade facilitation measures enhance the exports of goods from the middle-income countries but they may require improvement to exert stronger influence on the performance of the middle-income countries in terms of exports of goods.

The study also found that the lagged dependent variables is positive and statistically significant at influencing the exports of goods among the middle-income countries. The results show that number of instruments are at most (35) that are less than the number of groups (94). The Sargan and Hansen tests of over-identifying restrictions are valid as expected in supporting the choice of instruments, the AR (2) value of the models estimated are greater than the critical value (0.05) implying that the original error term is serially uncorrelated and the moment conditions are correctly specified. The study, using the Wald chi2 statistic, also found that there is a weak joint influence of all trade facilitation indicators on exports of goods.

### 4.6 *Impact of Trade Facilitation on Economic Growth among Middle-Income Countries*

The results from Table 6 explaining the impact of trade facilitation on economic growth among middle-income countries were estimated using the difference GMM and system GMM. The results show that container port traffic has a significant positive influence on economic growth among the middle-income countries at 10% level of significance from the one-step system GMM. The study also showed that the efficiency of the customs clearance process has a significant positive influence on economic growth among the middle-income countries at 5% level of significance. On the other hand, the study found a significant negative influence of quality of trade and transport-related infrastructure and tariff rate on economic growth among the middle-income countries at 5% level of significance from the two-step difference GMM and one-step system GMM estimates. The implication is that the level of infrastructure in the middle-income countries or developing countries has contributed to deteriorating their level of economic growth. The estimates also show that tariff is not a good policy tool used by developing countries as it exerts strong negative effects on economic growth of the countries. The other variables specified had weak influence on economic growth among the countries. The study findings also showed that gross fixed capital formation has positive influence on economic growth among the countries.

From the results in Table 6, the study also found that the lagged dependent variables is positive and statistically significant at influencing the exports of goods among the middle-income countries. The results show that number of instruments are at most (35) that are less than the number of groups (94). The Sargan and Hansen tests of over-identifying restrictions are valid as expected in supporting the choice of instruments, the AR (2) value of the models estimated are greater than the critical value (0.05) implying that the original error term is serially uncorrelated and the moment conditions are correctly specified. The study, using the Wald chi2 statistic, also found that there is weak joint influence of all trade facilitation indicators on exports of goods.

### 4.7 *Impact of Trade Facilitation on Economic Growth among Lower-middle-income Countries*

The results in Table 7 and Table 8 explain the influence of trade facilitation on goods exports, goods imports and economic growth among the middle-income countries taking into account of the income classification of the countries (That is, high middle-income and low-middle-income countries respectively). The results show that container port traffic has significant influence on import of goods among the middle-income countries. The study also found that foreign direct investment discourages both export of goods and import of goods significantly at 5% level of significance. This means that the foreign firms turn to compete with the domestic firms leading to reduction of productive investment and killing of infant industries that could accelerate economic growth among the middle-income countries. This suggests that foreign direct investment should be considered as supplementary investment to the domestic investment. In that way, it will boost the export of goods from the middle-income countries. The estimated influence of container port traffic as presented in Table 7 also shows strong influence on goods imports by the middle-income countries. According to the findings in Table 7, the lagged dependent variables have a positive and statistically significant influence on middle-income nations' exports of products. The results demonstrate that the number of instruments is limited to 35, which is less than the number of groups (94). The Sargan and Hansen tests of over-identifying restrictions support the choice of instruments as expected; the AR (2) value of the models estimated is greater than the critical value (0.05), implying that the original error term is serially uncorrelated and the moment conditions are correctly specified.

### *4.8 Impact of Trade Facilitation on Economic Growth among Upper-middle-income Countries*

From the results in Table 8, the estimated impact of quality of trade and transport-related infrastructure has statistical significant positive influence on goods exports and goods imports at 5% level of significance. This explains the role of technology in effective management of customs and cross borders. More so, the influence of the level of frequency with which shipments reach consignee within scheduled or expected time on export of goods and services. Other variables were not statistically significant in influencing export of goods and import of goods. According to the findings in Table 8, the lagged dependent variables have a positive and statistically significant influence on the product exports of middle-income countries. The findings show that the number of instruments is restricted to 35, which is less than the number of groups (94). The Sargan and Hansen over-identifying restriction tests validate the instrument choice as expected; the AR (2) value of the estimation procedure is greater than the crucial value (0.05), implying that the original error term is serially uncorrelated and the moment requirements are correctly defined.

## 5. Conclusion and Policy Recommendations

### 5.1 *Conclusion*

The study infers that trade facilitation influences the level of trade flows and economic growth among the middle-income countries, and that the upper-middle-income countries benefit from the the quality of trade and transport-related infrastructure and container port more than the low-middle countries.

### 5.2 *Policy Recommendations*

The study findings about trade facilitation have important policy implications for the middle-income countries. However, the prevalence of complex and cumbersome border procedural requirements and other forms of institutional trade costs in among countries inflate the costs of moving goods across borders. In turn, this erodes the competiveness of local firms in foreign markets from the middle-income countries. These institutional trade costs have been a major impediment to middle-income nations reaping the full economic benefits of opening up to international trade. Given the large potential gains from improved trade facilitation, this paper suggests that reforms aimed at significantly lowering the costs of trading across borders among middle-income countries should be highly prioritized in policy formulations, with a focus on the export side by reducing at-the-border documentation, time, and real costs of trading across borders.

Since many of the middle-income countries are poised to reap greater benefits from trade facilitation initiatives than their average counterparts in the developing world, they also tend to respond more on the export side than on the import side, suggesting a positive adjustment of the balance of payments, at least in the short run.

The international organizations should continue to report the set of Trade Facilitation Indicators (TFIs) that identify areas for action and enable the potential impact of reforms to be assessed. This will enable countries to plan effectively trade-related policies and regulations and their implementation in practice in order to improve their border procedures, reduce trade costs, boost trade flows and reap greater benefits from international trade. Governments should also undergo and improve their border procedures, reduce trade costs, boost trade flows and reap greater benefits from international trade. This can be done through identifying and understanding the areas for action and enable the potential impact of reforms.

**References**

Adam Smith Institute (2021). The Wealth of Nations. Available at https://www.adamsmith.org/the-wealth-of-nations on 20th November, 2021.

Ali, W. & Shakoor, N. (2020). The Impact of Trade Facilitation on Trade Flow in Asian Countries. *Journal of Business & Financial Affairs.* 9(4), 1-9.

Amoako-Tuffour, J., Balchin, N., Calabrese, L. & Mendez-Parra, M. (2016). Trade facilitation and economic transformation in Africa. ACET-ATF.

Anjande G., Ijirshar, V. U., Asom, S. T., Akiri, S. E. & Sokpo, J. T. (2020). Revisiting the accuracy of Ricardian theory of comparative advantage in Africa in the 21st century. *Journal of Economics and Allied Research,* 4(3): 25-44.

Arellano, M., & Bond, S. (1991). Some tests of specification for panel data: Monte Carlo evidence and an application to employment equations. *Review of Economic Studies,* 58: 277–297.

Arellano, M., & Bover, O. (1995). Another look at the instrumental variable estimation of error-components models. *Journal of Econometrics,* 68: 29–51.

Avetisyan, M. & Hertel, T. (2021). Impacts of trade facilitation on modal choice and international trade flows. *Economics of Transportation,* 28, *Journal of Economics and Development*, 21, 69-80

Baltagi, B. H. (2008). *Econometric analysis of panel data*. 6$^{th}$ Edition. Wiley. Chichester.

Blundell, R., &. Bond, S. (1998). Initial conditions and moment restrictions in dynamic panel data models. *Journal of Econometrics,* 87: 115–143.

Brueckner, M. & Lederman, D. (2015). Trade openness and economic growth: panel data evidence from Sub-Saharan Africa. *Economica*, 82, 1302–1323.

Bruecknera, P. K., Czerny, A. I. & Gaggeroc, A. A. (2021). Airline schedule buffers and flight delays: A discrete model. Economics of Transportation, 26–27, 1-7.

Dollar, D. & Kraay, A. (2004). Trade, growth, and poverty. *The Economic Journal*, 114(493), F22–F49.

Dumitrescu, E. J. & Hurlin, C. (2012). Testing for Granger non-Causality in heterogeneous panels. *Economic Modelling*, 29(4), 1450-1460.

Fincke, A. (2006). Factor Endowments and Trade II: The Heckscher-Ohlin Model (Chapter 6). 095-120

Florensa, L. M., Márquez-Ramos, L., & Recalde, M. L. (2015). The effect of economic integration and institutional quality of trade agreements on trade margins: evidence for Latin America. *Review of World Economics*, 151(2), 329–351.

Frederick, G. (2018). Eighteenth-century Scottish political economy and the decline of imperial Spain. *Journal of Scottish Historical Studies*, 38(1), 55-72.

Goldin, I. & Reinert, K. (2007). *Globalization for Development: Trade, Finance, Aid, and Migration and Policy*. Rev. Ed. New York: World Bank and Palgrave Macmillan.

Heckscher, E. (1949). The Effect of Foreign Trade on the Distribution of Income," in H. S. Ellis and L. A. Metzler (eds.). *Readings in the Theory of International Trade*. Homewood, Ill.: Richard D. Irwin, Inc.

Herzer, D. (2013). Cross-country Heterogeneity and the trade-income relationship. *World Development*, 44, 194–211.

Heshmati, A. & Peng, S. (2012). International trade and its effects on economic performance in China. *China Economic Policy Review*, 1(2), 35–61.

Hill, C. W. L & Arun, K. J. (2014). *Internationl Business: Competing in the Global Marketplace.* India: MMcGraw Hill Education Private Limited.

Ijirshar, V. U. (2019). Impact of trade openness on economic growth among ECOWAS countries. *CBN Journal of Applied Statistics*, 10(1), 75-96.

Internal Trade Centre (2021). Trade Facilitation. Available at https://www.intracen.org/itc/policy/trade-facilitation/ on 8th November, 2021.

International Chamber of Commerce (2021). Customs & trade facilitation. Available at https://iccwbo.org/global-issues-trends/trade-investment/trade-facilitation/ on 20th November 2021.

Irarrazabal, A.., Moxnes, A. & Opromolla, L. D. (2010). The tip of the iceberg: modeling trade costs and implications for intraindustry reallocation. EFIGE working paper 22 and CEPR Discussion Paper Series No.7685

Kouty, M, (2021). Implementing the African Continental Free Trade Area (AfCFTA): The effects of trade procedures on trade flows. *Research in Applied Economics*, 13(1), 15-31.

Krugman, P. (1979). Increasing Returns, Monopolistic Competition, and International Trade. Journal of International Economics, 9, 469-480.

Krugman, P. (1980). Scale Economies, Product Differentiation, and the Pattern of Trade. *The American Economic Review, 70*(5), 950–959.

Kummer-Noormamode, S. (2014). Does trade with china have an impact on African countries' growth? *African Development Review*, 26(2), 397–415.

Ma´rquez-Ramos, L., & Martinez-Gomez, V. (2014). A gravity assessment of Moroccan F&V monthly exports to EU countries: The effect of trade preferences revisited. Poster paper prepared for presentation at the EAAE 2014 Congress 'Agri-Food and Rural Innovations for Healthier Societies' August 26–29, Ljubljana, Slovenia.

Mai, N. T. T. & Ngoc, N. B. (2019). Trade Facilitation in Vietnam: Estimating The Effects on Trade Flows. National Economics University, Vietnam

Melitz, M. J. (2003). The Impact of Trade on Intra-Industry Reallocations and Aggregate Industry Productivity. *Econometrica*, *71*(6), 1695–1725.

Mun, T. (1664). *Mercantilism*. Available at https://teachingamericanhistory.org/document/mercantilism/ on 20th November, 2021.

Odularu, G. & Alege, P. (2019). *Trade Facilitation Capacity Needs (Policy Directions for National and Regional Development in West Africa)* || Bilateral Trade Flows, Trade Facilitation, and RTAs: Lessons from ECOWAS. (Chapter 4), 67–90.

Odularu, G. & Alege, P. (2019). Trade facilitation capacity needs (policy directions for national and regional development in West Africa) || *Bilateral Trade Flows, Trade Facilitation, and RTAs: Lessons from ECOWAS*. (Chapter 4), 67–90.

Ohlin, B. (1933). *Interregional and International Trade*. Cambridge, Mass.: Harvard Univ. Press.

Olubusoye, E. O., Salisu, A. A. & Olofin, S. O. (2016). *Applied panel data analysis*. Centre for Econometric and Allied Research (CEAR), University of Ibadan, Ibadan.

Perera, S., Siriwardana, M. & Mounter, S. (2017). Trade Facilitation, Economic Development and Poverty Alleviation: South Asia at a Glance. In G. I. Staicu, Poverty, Inequality and Policy. (Chapter 7).

Rehim K. C. (2002). Lecture Notes: Absolute and comparative advantage: Ricardian model. Department of Economics, Michigan State University, East Lansin.

Ricardo, D. (1817). Principles of political economy and taxation, reprinted by J.M. Dent, London, in Everyman's Library, 1911.

Sá Porto, P. C. D., Canuto, O., & Morini, C. (2015). The impacts of trade facilitation measures on international trade flows. *World Bank Policy Research Working Paper*, (7367).

Safaeimanesh, S., & Jenkins, G. P. (2021). Trade facilitation and its impacts on the economic welfare and sustainable development of the ECOWAS region. *Sustainability,* 13(1), 164, 1-22.

Sakyi, D., Villaverde, J. & Maza, A. (2015). Trade openness, income levels, and economic growth: the case of developing countries, 1970–2009. *The Journal of International Trade & Economic Development*, 24(6), 860–82.

Sakyi, D., Villaverde, J., Maza, A., & Bonuedi, I. (2017). The effects of trade and trade facilitation on economic growth in Africa. *African Development Review*, 29(2), 350–361.

Samuelson P. A. (1954). The transfer problem and transport costs, II: Analysis of effects of trade impediments. *The Economic Journal*, 64(254), 264-289.

Seck, A. (2016). Trade facilitation and trade participation: Are sub-Saharan African firms different? *Journal of African Trade*, 3(1-2), 23–39.

Seetanah, B. Sannassee, R. & Fauzel, S. (2016), trade facilitation and trade flows: evidence from Africa", in Teh, R., et al. (eds.), *Trade Costs and Inclusive Growth: Case Studies Presented by WTO Chair-Holders*, WTO, Geneva, https://doi.org/10.30875/49648549-en.

Solow, R. M. (1956). A contribution to the theory of economic growth. Quarterly Journal of Economics Oxford Journals, 70(1), 65–94.

Swan, T. W. (1956). Economic growth and capital accumulation. Economic Record. Wiley, 32(2), 334–361.

Thompson, H. (2017). International Economics Global Markets and Competition. Constant Cost Production and Trade (Chapter 5). Available at https://www.worldscientific.com/doi/pdf/10.1142/9789814663885_0005 on 20th November, 2021.

United Nations (2016). Trade Facilitation and Development: Driving trade competitiveness, border agency effectiveness and strengthened governance. United Nations Conference on Trade and Development UNCTAD/DTL/TLB/2016/1. Available at https://unctad.org/system/files/official-document/dtltlb2016d1_en.pdf on 8th November, 2021.

United Nations (2021). Trade facilitation - principles and benefits. *Trade Facilitation Guide*. Available at https://tfig.unece.org/details.html on 8th November, 2021.

United Nations Conference on Trade and Development (UNCTAD) (2021). Trade facilitation. Available at https://unctad.org/topic/transport-and-trade-logistics/trade-facilitation on 8th November, 2021.

Wangwe, S. (1993). New Trade Theories and Developing Countries: Policy and Technological Implications. The United Nations University, Working Paper No. 7.

Wilson, J. S., Mann, C. L., & Otsuki, T. (2003). Trade facilitation and economic development: a new approach to quantifying the impact. *The World Bank Economic Review*, *17*(3), 367–389. http://www.jstor.org/stable/3990246

World Bank (2018). Stronger Open Trade Policies Enable Economic Growth for All. Available at https://www.worldbank.org/en/results/2018/04/03/stronger-open-trade-policies-enables-economic-growth-for-all on 20th November, 2021.

World Bank (2021). *Doing Business 2020 Comparing Business Regulation in 190 Economies 2020*. Doing Business provides objective measures of business regulations for local firms in 190 economies. International Bank for Reconstruction and Development. Available at https://www.doingbusiness.org/en/doingbusiness on 20th November, 2021.

World Bank (2021). *The World Bank in Middle-income Countries*. Available at https://www.worldbank.org/en/country/mic/overview on 8th November, 2021.

World Trade Organisation (WTO) (2021). Trade Facilitation Indicators.

World Trade Organization (2017). World Trade Report 2015. Available from: https://www.wto.org/english/res_e/booksp_e/world_trade_report15_e.pdf [cited: 3 February 2017]

YI, K.-M. (2003). Can Vertical Specialization Explain the Growth of World Trade? *Journal of Political Economy*, 111, 52–102.

Yu, Z. & Luu, M. B. (2020). The impact of trade facilitation on Vietnam's trade flows. *ASEAN Journal of Management & Innovation*, 7(1), 133 – 152.

**Table 1: Variable Description and Measurement**

| Label | Variable | Definition | Measurement | Source |
|---|---|---|---|---|
| GNI | Gross National Income, PPP | Gross national income is the sum of value added by all resident producers plus any product taxes (less subsidies) plus net receipts of primary income from abroad. It is per capita values for gross national income expressed in current international dollars converted by Purchasing Power Parity (PPP) conversion factor. | Current International Dollars. | The World Bank |
| EXPG | Goods Exports | Goods exports refer to all movable goods involved in a change of ownership from residents to nonresidents. | Current U.S. Dollars. | The World Bank |
| IMPG | Goods imports | Goods imports refer to all movable goods involved in a change of ownership from nonresidents to residents. | Current U.S. Dollars. | The World Bank |
| LPIAC | Ability to track and trace consignments | This is the ability to track and trace consignments when shipping to the market, on a rating ranging from 1 (very low) to 5 (very high). | Index (1=low to 5=high) | World Bank |
| LPICQ | Competence and quality of logistics services | This measure the overall level of competence and quality of logistics services on a rating ranging from 1 (very low) to 5 (very high). | Index (1=low to 5=high) | World Bank |
| LPIEA | Ease of arranging competitively priced shipments | This is the ease of arranging competitively priced shipments to markets, on a rating ranging from 1 (very difficult) to 5 (very easy). | Index (1=low to 5=high) | World Bank |
| LPIEC | Efficiency of customs clearance process | This measures efficiency of customs clearance processes on a rating ranging from 1 (very low) to 5 (very high). | Index (1=low to 5=high) | World Bank |
| LPIFS | Frequency with which shipments reach consignee within scheduled or expected time | This assesses how often the shipments to assessed markets reach the consignee within the scheduled or expected delivery time, on a rating ranging from 1 (hardly ever) to 5 (nearly always). | Index (1=low to 5=high) | World Bank |
| LPI | Logistics performance index, Overall | This evaluates eight markets on six core dimensions on a scale from 1 (worst) to 5 (best). | Index (1=low to 5=high) | World Bank |
| PCT | Port container traffic | Port container traffic measures the flow of containers from land to sea transport modes, and vice versa, in Twenty-Foot Equivalent Units (TEUs), a standard-size container. | 20 foot equivalent units | World Bank |
| ATCE | Average time to clear exports through customs | Average time to clear exports through customs is the average number of days to clear direct exports through customs. | Days | World Bank |
| AFT | Air freight transportation | Air freight is the volume of freight, express, and diplomatic bags carried on each flight stage measured in metric tons times kilometers traveled. | million ton-km | World Bank |
| TRF | Tariff rate (applied, weighted mean) | Weighted mean applied tariff is the average of effectively applied rates weighted by the product import shares corresponding to each partner country. | weighted mean | World Bank |
| QPI | Quality of Port Infrastructure | The Quality of Port Infrastructure measures business executives' perception of their country's port facilities | Index (1=Extremely Underdeveloped To 7=Well Developed) | World Economic Forum Global Competitiveness Report |
| FDI | Foreign direct investment | Foreign direct investment are the net inflows of investment to acquire a lasting management interest in an enterprise operating in an economy other than that of the investor. | Current U.S. Dollars | World Bank |
| GFCF | Gross fixed capital formation | Gross fixed capital formation includes land improvements; plant, machinery, and equipment purchases; and the construction of roads, railways, and the like, including schools, offices, hospitals, private residential dwellings, and commercial and industrial buildings. | Current U.S. Dollars | World Bank |

**Source: Author's Compilations**

**Table 2: Descriptive Statistics**

| | Middle-income Countries (Obs=1034) | | | | Upper-middle-income Countries (Obs=506) | | | | Lower-middle-income Countries (Obs=528) | | | |
|---|---|---|---|---|---|---|---|---|---|---|---|---|
| Variable | Mean | Std | Min | Max | Mean | Std | Min | Max | Mean | Std | Min | Max |
| GNI | 10179.27 | 6256.525 | 1610 | 31840 | 14855.94 | 5208.089 | 4500.667 | 31840 | 5697.473 | 3100.357 | 1610 | 18010 |
| EXPG | 6.32e+10 | 2.30e+11 | 1.09e+07 | 2.50e+12 | 9.97e+10 | 3.19e+11 | 1.98e+08 | 2.50e+12 | 2.82e+10 | 6.35e+10 | 1.09e+07 | 7.42e+11 |
| IMPG | 5.99e+10 | 1.91e+11 | 9.62e+07 | 2.04e+12 | 8.79e+10 | 2.59e+11 | 1.07e+09 | 2.04e+12 | 3.31e+10 | 7.55e+10 | 9.62e+07 | 6.23e+11 |
| ATCE | 7.829449 | 4.358968 | 1 | 26 | 7.512621 | 4.40467 | 1 | 21.7 | 8.133076 | 4.296988 | 1.8 | 26 |
| PCT | 4716163 | 1.96e+07 | 472.614 | 2.42e+08 | 6473584 | 2.72e+07 | 472.614 | 2.42e+08 | 3031969 | 6157906 | 3303.152 | 6.62e+07 |
| AFT | 529.5171 | 2166.075 | 0 | 25394.59 | 844.2934 | 2993.477 | 0 | 25394.59 | 227.8565 | 650.8994 | 0 | 8596.939 |
| QPI | 3.567157 | .9346034 | 1.3 | 6.4 | 3.74705 | .9403937 | 1.61 | 6.4 | 3.394759 | .8965979 | 1.3 | 5.5 |
| LPIAC | 2.654522 | .3751715 | 1.542857 | 3.915447 | 2.721116 | .4016119 | 1.64 | 3.915447 | 2.590702 | .3361563 | 1.542857 | 3.518981 |
| LPICQ | 2.571175 | .3461395 | 1.681079 | 3.747822 | 2.640872 | .3747325 | 1.75 | 3.747822 | 2.504383 | .3019186 | 1.681079 | 3.4 |
| LPIEA | 2.67699 | .3501937 | 1.77 | 3.704961 | 2.742434 | .3500987 | 1.890625 | 3.704961 | 2.614273 | .3389381 | 1.77 | 3.438571 |
| LPIEC | 2.411796 | .3107561 | 1.5 | 3.59546 | 2.470341 | .3305277 | 1.68 | 3.59546 | 2.355691 | .2795677 | 1.5 | 3.17442 |
| LPIFS | 3.10056 | .3705778 | 2.023799 | 4.14 | 3.175554 | .3773129 | 2.308838 | 4.14 | 3.02869 | .3495521 | 2.023799 | 4 |
| LPI | 2.649708 | .3203327 | 1.716096 | 3.775321 | 2.722053 | .3428095 | 1.87919 | 3.775321 | 2.580377 | .280506 | 1.716096 | 3.420043 |
| LPIQTT | 2.447412 | .3735604 | 1.27 | 3.79 | 2.551193 | .4042254 | 1.5 | 3.79 | 2.347956 | .3109751 | 1.27 | 3.337178 |
| TRF | 6.278156 | 4.355318 | .3 | 35.65 | 5.090698 | 3.84181 | .3 | 23.97 | 7.416136 | 4.5152 | .73 | 35.65 |
| FDI | -4.07e+09 | 1.68e+10 | -2.32e+11 | 4.17e+10 | -5.84e+09 | 2.29e+10 | -2.32e+11 | 4.17e+10 | -2.38e+09 | 6.62e+09 | -9.33e+10 | 8.75e+09 |
| GFCF | 8.99e+10 | 4.69e+11 | 1.26e+08 | 6.12e+12 | 1.41e+11 | 6.59e+11 | 4.99e+08 | 6.12e+12 | 4.12e+10 | 1.06e+11 | 1.26e+08 | 8.25e+11 |

Source: Extractions from STATA 15 Output

Table 3: Dumitrescu & Hurlin Granger non-causality test results

| Null Hypothesis | W-bar | Z-bar | Prob | Z-bar tilde | Prob | Decision |
|---|---|---|---|---|---|---|
| lnEXPG does not Granger-cause lnGNI | 2.7476 | 11.9806 | 0.0000 | 4.6661 | 0.0000 | Reject Ho |
| lnGNI does not Granger-cause lnEXPG. | 4 .0522 | 20.9252 | 0.0000 | 9.1838 | 0.0000 | Reject Ho |
| lnIMPG does not Granger-cause lnGNI. | 3.7233 | 18.6701 | 0.0000 | 8.0448 | 0.0000 | Reject Ho |
| lnGNI does not Granger-cause lnIMPG | 5.3361 | 29.7271 | 0.0000 | 13.6294 | 0.0000 | Reject Ho |
| lnGNI does not Granger-cause ATCE | 46.5882 | 312.5372 | 0.0000 | 156.4701 | 0.0000 | Reject Ho |
| lnPCT does not Granger-cause lnGNI | 1.9573 | 6.5627 | 0.0000 | 1.9296 | 0.0537 | Reject Ho |
| lnGNI does not Granger-cause AFT | 3.3399 | 16.0415 | 0.0000 | 6.7171 | 0.0000 | Reject Ho |
| lnGNI does not Granger-cause QPI | 2.1001 | 7.5422 | 0.0000 | 2.4243 | 0.0153 | Reject Ho |
| lnGNI does not Granger-cause LPIAC | 1.8239 | 5.6486 | 0.0000 | 1.4679 | 0.1421 | Reject Ho |
| lnGNI does not Granger-cause LPICQ | 2.1136 | 7.6343 | 0.0000 | 2.4708 | 0.0135 | Reject Ho |
| lnGNI does not Granger-cause LPIEA. | 2.1627 | 7.9714 | 0.0000 | 2.6411 | 0.0083 | Reject Ho |
| lnGNI does not Granger-cause LPIFS | 2.3620 | 9.3374 | 0.0000 | 3.3310 | 0.0009 | Reject Ho |
| lnGNI does not Granger-cause LPI. | 2.2265 | 8.4085 | 0.0000 | 2.8619 | 0.0042 | Reject Ho |
| lnGNI does not Granger-cause LPIQTT | 3.0032 | 13.7334 | 0.0000 | 5.5514 | 0.0000 | Reject Ho |
| lnGNI does not Granger-cause TRF. | 2.1315 | 7.7571 | 0.0000 | 2.5329 | 0.0113 | Reject Ho |
| lnIMPG does not Granger-cause lnEXPG | 1.54e+06 | 1.06e+07 | 0.0000 | 5.34e+06 | 0.0000 | Reject Ho |
| lnEXPG does not Granger-cause lnIMPG | 350.7482 | 2397.7528 | 0.0000 | 1209.6630 | 0.0000 | Reject Ho |
| lnEXPG does not Granger-cause ATCE | 10.6980 | 66.4862 | 0.0000 | 32.1956 | 0.0000 | Reject Ho |
| lnPCT does not Granger-cause lnEXPG. | 12.0508 | 75.7608 | 0.0000 | 36.8799 | 0.0000 | Reject Ho |
| lnEXPG does not Granger-cause AFT | 4.3202 | 22.7624 | 0.0000 | 10.1117 | 0.0000 | Reject Ho |
| lnEXPG does not Granger-cause QPI. | 6.4331 | 37.2474 | 0.0000 | 17.4277 | 0.0000 | Reject Ho |
| lnEXPG does not Granger-cause LPIAC. | 2.4183 | 9.7231 | 0.0000 | 3.5258 | 0.0004 | Reject Ho |
| lnEXPG does not Granger-cause LPICQ | 3.1185 | 14.5236 | 0.0000 | 5.9505 | 0.0000 | Reject Ho |
| lnEXPG does not Granger-cause LPIEA. | 1.9943 | 6.8163 | 0.0000 | 2.0577 | 0.0396 | Reject Ho |
| lnEXPG does not Granger-cause LPIEC. | 3.0693 | 14.1867 | 0.0000 | 5.7803 | 0.0000 | Reject Ho |
| lnEXPG does not Granger-cause LPIFS | 2.9589 | 13.4294 | 0.0000 | 5.3978 | 0.0000 | Reject Ho |
| lnEXPG does not Granger-cause LPI | 3.9526 | 20.2422 | 0.0000 | 8.8388 | 0.0000 | Reject Ho |
| lnEXPG does not Granger-cause LPIQTT | 2.9087 | 13.085 | 0.0000 | 5.2240 | 0.0000 | Reject Ho |
| lnEXPG does not Granger-cause TRF. | 6.2246 | 35.8180 | 0.0000 | 16.7058 | 0.0000 | Reject Ho |
| lnIMPG does not Granger-cause ATCE | 9.8244 | 60.4972 | 0.0000 | 29.1707 | 0.0000 | Reject Ho |
| lnPCT does not Granger-cause lnIMPG | 13.7706 | 87.5505 | 0.0000 | 42.8346 | 0.0000 | Reject Ho |
| lnIMPG does not Granger-cause AFT | 4.7636 | 25.8017 | 0.0000 | 11.6468 | 0.0000 | Reject Ho |
| lnIMPG does not Granger-cause QPI. | 6.5625 | 38.1345 | 0.0000 | 17.8758 | 0.0000 | Reject Ho |
| lnIMPG does not Granger-cause LPIAC | 2.0094 | 6.9198 | 0.0000 | 2.1100 | 0.0349 | Reject Ho |
| lnIMPG does not Granger-cause LPICQ | 3.1585 | 14.7980 | 0.0000 | 6.0891 | 0.0000 | Reject Ho |
| lnIMPG does not Granger-cause LPIEA | 2.4373 | 9.8537 | 0.0000 | 3.5918 | 0.0003 | Reject Ho |
| lnIMPG does not Granger-cause LPIEC | 2.9660 | 13.4784 | 0.0000 | 5.4226 | 0.0000 | Reject Ho |
| lnIMPG does not Granger-cause LPIFS | 2.6698 | 11.4474 | 0.0000 | 4.3968 | 0.0000 | Reject Ho |
| lnIMPG does not Granger-cause LPI | 3.7781 | 19.0458 | 0.0000 | 8.2345 | 0.0000 | Reject Ho |
| lnIMPG does not Granger-cause LPIQTT | 2.7380 | 11.9152 | 0.0000 | 4.6330 | 0.0000 | Reject Ho |
| lnIMPG does not Granger-cause LPIQTT | 5.8996 | 33.5896 | 0.0000 | 15.5803 | 0.0000 | Reject Ho |

Source: Extractions from STATA 15 Output

Table 4: Impact of Trade Facilitation on Export Goods among Middle-Income Countries

| | With the Effects of Logistic Performance Indices | | | | With the Effect of Overall Logistic Performance Index | | | |
|---|---|---|---|---|---|---|---|---|
| | (1) | (2) | (3) | (4) | (5) | (6) | (7) | (8) |
| VARIABLES | One-step Difference GMM | Two-step Difference GMM | One-step System GMM | Two-step System GMM | One-step Difference GMM | Two-step Difference GMM | One-step System GMM | Two-step System GMM |
| | lnEXPG | lnEXPG | lnEXPG | lnEXPG | lnEXPG | lnEXPG | lnEXPG | lnEXPG |
| L.lnEXPG | 1.229*** | 1.104*** | 0.741*** | 0.802*** | 1.212*** | 1.109*** | 0.744*** | 0.820*** |
| | (0.277) | (0.261) | (0.173) | (0.158) | (0.271) | (0.262) | (0.183) | (0.175) |
| ATCE | -0.0143 | -0.0168 | -0.00436 | -0.00355 | -0.0178 | -0.0185 | -0.00680 | -0.00349 |
| | (0.0124) | (0.0140) | (0.00625) | (0.00519) | (0.0118) | (0.0127) | (0.00704) | (0.00515) |
| LnPCT | 0.128** | 0.108 | 0.0948 | 0.0648* | 0.124** | 0.103 | 0.0949 | 0.0589 |
| | (0.0603) | (0.0720) | (0.0571) | (0.0362) | (0.0598) | (0.0711) | (0.0591) | (0.0376) |
| AFT | 0.000139 | 0.000113 | 1.29e-06 | 2.09e-05 | 0.000139 | 0.000114 | 1.26e-06 | 2.02e-05 |
| | (0.000102) | (0.000139) | (2.31e-05) | (2.16e-05) | (0.000103) | (0.000147) | (2.49e-05) | (2.36e-05) |
| QPI | -0.0402 | -0.0337 | -0.0125 | -0.0133 | -0.0402 | -0.0383 | -0.0144 | -0.0169 |
| | (0.0463) | (0.0429) | (0.0314) | (0.0291) | (0.0471) | (0.0422) | (0.0329) | (0.0272) |
| LPIAC | -0.210* | -0.0594 | -0.153 | -0.0257 | | | | |
| | (0.112) | (0.102) | (0.120) | (0.0991) | | | | |
| LPICQ | 0.362** | 0.176 | 0.183 | 0.142 | | | | |
| | (0.159) | (0.125) | (0.143) | (0.163) | | | | |
| LPIEA | 0.0193 | -0.00563 | 0.0142 | 0.122 | | | | |
| | (0.0714) | (0.0607) | (0.155) | (0.198) | | | | |
| LPIEC | -0.104 | -0.0647 | -0.235 | -0.249* | | | | |
| | (0.114) | (0.107) | (0.153) | (0.132) | | | | |
| LPIFS | 0.176 | 0.0523 | 0.202 | 0.0799 | | | | |
| | (0.112) | (0.0877) | (0.145) | (0.110) | | | | |
| LPIQTT | 0.00915 | 0.151 | 0.133 | 0.198** | | | | |
| | (0.158) | (0.122) | (0.0981) | (0.0944) | | | | |
| TRF | 0.0111 | 0.00371 | -0.00989 | -0.00856 | 0.0134 | 0.00530 | -0.00944 | -0.00781 |
| | (0.00798) | (0.00635) | (0.0101) | (0.0124) | (0.00820) | (0.00657) | (0.0113) | (0.0137) |
| lnFDI | -0.439 | -0.318 | -0.126 | -0.148 | -0.435 | -0.321 | -0.121 | -0.126 |
| | (0.323) | (0.435) | (0.138) | (0.152) | (0.326) | (0.465) | (0.139) | (0.143) |
| lnGFCF | 0.0464 | -0.0368 | 0.141 | 0.0656 | 0.0579 | -0.0323 | 0.148 | 0.0649 |
| | (0.0958) | (0.0819) | (0.147) | (0.138) | (0.0966) | (0.0775) | (0.158) | (0.159) |
| yr1 | 0 | 0 | 0 | 0 | 0 | 0 | 0 | 0 |
| | (0) | (0) | (0) | (0) | (0) | (0) | (0) | (0) |
| yr2 | 0.195*** | 0.184** | 4.466 | 0.0512 | 0.159 | 0 | 4.264 | 0 |
| | (0.0694) | (0.0865) | (3.875) | (0.0853) | (0.0999) | (0) | (3.888) | (0) |
| yr3 | 0 | -0.0362 | 4.360 | -0.102 | -0.0392 | -0.214*** | 4.146 | -0.163*** |
| | (0) | (0.0746) | (3.885) | (0.109) | (0.0788) | (0.0664) | (3.895) | (0.0420) |
| yr4 | -0.0215 | -0.0506 | 4.362 | -0.111 | -0.0639 | -0.231*** | 4.139 | -0.175*** |
| | (0.0460) | (0.0776) | (3.884) | (0.101) | (0.0845) | (0.0675) | (3.893) | (0.0308) |
| yr5 | -0.0656** | -0.0653 | 4.335 | -0.125 | -0.111 | -0.249*** | 4.102 | -0.191*** |
| | (0.0325) | (0.0776) | (3.886) | (0.0974) | (0.0811) | (0.0669) | (3.895) | (0.0266) |
| yr6 | -0.242*** | -0.236*** | 4.163 | -0.277** | -0.282*** | -0.418*** | 3.934 | -0.344*** |
| | (0.0330) | (0.0798) | (3.883) | (0.106) | (0.0811) | (0.0760) | (3.893) | (0.0442) |
| yr7 | -0.0394 | -0.0732 | 4.278 | -0.156 | -0.0784 | -0.255*** | 4.055 | -0.225*** |
| | (0.0614) | (0.0780) | (3.883) | (0.0947) | (0.0881) | (0.0316) | (3.894) | (0.0244) |
| yr8 | 0.0967* | 0.0546 | 4.409 | -0.0419 | 0.0502 | -0.128*** | 4.183 | -0.113*** |
| | (0.0512) | (0.0844) | (3.880) | (0.0947) | (0.0921) | (0.0390) | (3.889) | (0.0234) |
| yr9 | 0.0337 | 0.0112 | 4.407 | -0.0533 | -0.0170 | -0.174*** | 4.177 | -0.128*** |
| | (0.0371) | (0.0834) | (3.882) | (0.0952) | (0.0872) | (0.0618) | (3.891) | (0.0211) |
| yr10 | -0.0832** | -0.0753 | 4.338 | -0.118 | -0.126 | -0.265*** | 4.120 | -0.188*** |
| | (0.0401) | (0.0880) | (3.882) | (0.0947) | (0.0905) | (0.0862) | (3.893) | (0.0423) |
| yr11 | 0.0321 | 0 | 4.466 | 0 | 0 | -0.196** | 4.260 | -0.0660 |
| | (0.0746) | (0) | (3.907) | (0) | (0) | (0.0888) | (3.918) | (0.0843) |
| LPI | | | | | 0.227* | 0.259* | 0.147 | 0.287*** |
| | | | | | (0.115) | (0.133) | (0.128) | (0.100) |
| Constant | | | 0 | 5.511 | | | 0 | 4.683 |
| | | | (0) | (4.240) | | | (0) | (4.009) |
| Observations | 846 | 846 | 940 | 940 | 846 | 846 | 940 | 940 |
| Number of country | 94 | 94 | 94 | 94 | 94 | 94 | 94 | 94 |
| No of Instruments | 31 | 31 | 33 | 33 | 26 | 26 | 28 | 28 |

Source: Extractions from STATA 15 Output

Table 5: Impact of Trade Facilitation on Import Goods among Middle-Income Countries

| | With the Effects of Logistic Performance Indices | | | | With the Effect of Overall Logistic Performance Index | | | |
|---|---|---|---|---|---|---|---|---|
| | (1) | (2) | (3) | (4) | (5) | (6) | (7) | (8) |
| VARIABLES | One-step Difference GMM lnIMPG | Two-step Difference GMM lnIMPG | One-step System GMM lnIMPG | Two-step System GMM lnIMPG | One-step Difference GMM lnIMPG | Two-step Difference GMM lnIMPG | One-step System GMM lnIMPG | Two-step System GMM LnIMPG |
| L.lnIMPG | 1.250*** | 1.156*** | 0.622*** | 0.791*** | 1.244*** | 1.156*** | 0.570*** | 0.746*** |
| | (0.367) | (0.361) | (0.188) | (0.200) | (0.362) | (0.334) | (0.198) | (0.200) |
| ATCE | -0.0178 | -0.0195 | -0.0148 | -0.00894 | -0.0194* | -0.0201 | -0.0179* | -0.0120 |
| | (0.0108) | (0.0121) | (0.00940) | (0.0103) | (0.0109) | (0.0123) | (0.0108) | (0.0115) |
| lnPCT | 0.112** | 0.0802 | 0.0936** | 0.0699 | 0.110** | 0.0832 | 0.106** | 0.0890* |
| | (0.0543) | (0.0562) | (0.0463) | (0.0469) | (0.0536) | (0.0564) | (0.0504) | (0.0523) |
| AFT | 0.000143 | 7.98e-05 | 5.59e-06 | -3.69e-06 | 0.000143 | 7.11e-05 | 4.05e-06 | -1.76e-06 |
| | (9.86e-05) | (0.000124) | (1.74e-05) | (1.22e-05) | (9.86e-05) | (0.000118) | (1.81e-05) | (1.29e-05) |
| QPI | -0.0382 | -0.0523 | -0.00532 | -0.0201 | -0.0371 | -0.0547 | -0.00952 | -0.0273 |
| | (0.0372) | (0.0341) | (0.0304) | (0.0303) | (0.0373) | (0.0333) | (0.0342) | (0.0323) |
| LPIAC | -0.0271 | 0.0631 | 0.0314 | -0.0411 | | | | |
| | (0.0733) | (0.0683) | (0.122) | (0.111) | | | | |
| LPICQ | 0.190 | 0.101 | 0.0499 | 0.129 | | | | |
| | (0.121) | (0.116) | (0.135) | (0.138) | | | | |
| LPIEA | -0.0112 | -0.0173 | 0.142 | 0.0340 | | | | |
| | (0.0431) | (0.0470) | (0.127) | (0.108) | | | | |
| LPIEC | -0.00220 | -0.0329 | -0.287* | -0.172 | | | | |
| | (0.105) | (0.106) | (0.151) | (0.116) | | | | |
| LPIFS | 0.107* | 0.0743 | 0.112 | 0.0976 | | | | |
| | (0.0615) | (0.0599) | (0.0872) | (0.0888) | | | | |
| LPIQTT | 0.0103 | 0.0642 | 0.225** | 0.177** | | | | |
| | (0.110) | (0.0926) | (0.0914) | (0.0785) | | | | |
| TRF | 0.00542 | 0.00267 | -0.0169* | -0.0106 | 0.00627 | 0.00306 | -0.0199* | -0.0132 |
| | (0.00644) | (0.00645) | (0.00881) | (0.00762) | (0.00654) | (0.00639) | (0.0104) | (0.00805) |
| lnFDI | -0.435 | -0.227 | -0.155 | -0.0666 | -0.433 | -0.200 | -0.158 | -0.113 |
| | (0.308) | (0.372) | (0.144) | (0.0933) | (0.307) | (0.350) | (0.152) | (0.112) |
| lnGFCF | 0.0292 | 0.0120 | 0.172 | 0.0736 | 0.0345 | 0.0216 | 0.211 | 0.0934 |
| | (0.0621) | (0.0704) | (0.125) | (0.136) | (0.0629) | (0.0717) | (0.136) | (0.133) |
| yr1 | 0 | 0 | 0 | 0 | 0 | 0 | 0 | 0 |
| | (0) | (0) | (0) | (0) | (0) | (0) | (0) | (0) |
| yr2 | 0.179** | 0.241** | 0.216*** | 3.685 | 0.179** | 0 | 0.109** | 0 |
| | (0.0795) | (0.117) | (0.0623) | (3.115) | (0.0796) | (0) | (0.0474) | (0) |
| yr3 | 0 | 0.0475 | 0.155*** | 3.557 | 0 | -0.196** | 0.0457** | -0.122*** |
| | (0) | (0.0653) | (0.0329) | (3.138) | (0) | (0.0791) | (0.0228) | (0.0395) |
| yr4 | -0.0300 | 0.0162 | 0.158*** | 3.542 | -0.0323 | -0.230*** | 0.0435** | -0.143*** |
| | (0.0347) | (0.0616) | (0.0277) | (3.137) | (0.0355) | (0.0865) | (0.0212) | (0.0345) |
| yr5 | -0.0945*** | -0.0164 | 0.119*** | 3.518 | -0.0987*** | -0.264*** | 0 | -0.171*** |
| | (0.0355) | (0.0611) | (0.0172) | (3.137) | (0.0348) | (0.0917) | (0) | (0.0327) |
| yr6 | -0.229*** | -0.142** | 0 | 3.402 | -0.232*** | -0.391*** | -0.119*** | -0.283*** |
| | (0.0413) | (0.0598) | (0) | (3.145) | (0.0399) | (0.0971) | (0.0145) | (0.0416) |
| yr7 | -0.110*** | -0.0447 | 0.0412 | 3.466 | -0.112*** | -0.293*** | - | -0.226*** |
| | | | | | | | 0.0846*** | |
| | (0.0369) | (0.0741) | (0.0254) | (3.135) | (0.0352) | (0.0605) | (0.0245) | (0.0312) |
| yr8 | 0.0480 | 0.104 | 0.162*** | 3.596 | 0.0443 | -0.144*** | 0.0316 | -0.103*** |
| | (0.0470) | (0.0867) | (0.0454) | (3.122) | (0.0462) | (0.0486) | (0.0403) | (0.0258) |
| yr9 | 0.0114 | 0.0784 | 0.186*** | 3.612 | 0.00704 | -0.167** | 0.0600** | -<br>0.0852*** |
| | (0.0285) | (0.0693) | (0.0380) | (3.128) | (0.0283) | (0.0745) | (0.0292) | (0.0284) |
| yr10 | -0.104** | -0.0342 | 0.133*** | 3.530 | -0.105** | -0.280** | 0.0242 | -0.158*** |
| | (0.0490) | (0.0713) | (0.0357) | (3.138) | (0.0470) | (0.110) | (0.0265) | (0.0512) |
| yr11 | -0.0432 | 0 | 0.203*** | 3.568 | -0.0409 | -0.239** | 0.104 | -0.122 |
| | (0.0710) | (0) | (0.0740) | (3.148) | (0.0700) | (0.108) | (0.0742) | (0.0862) |
| LPI | | | | | 0.258*** | 0.266*** | 0.316** | 0.271** |
| | | | | | (0.0816) | (0.0902) | (0.138) | (0.128) |
| Constant | | | 6.957 | 0 | | | 7.260 | 5.195 |
| | | | (4.485) | (0) | | | (4.672) | (3.597) |
| Observations | 846 | 846 | 940 | 940 | 846 | 846 | 940 | 940 |
| Number of country | 94 | 94 | 94 | 94 | 94 | 94 | 94 | 94 |
| No of Instruments | 31 | 31 | 33 | 33 | 26 | 26 | 28 | 28 |

Source: Extractions from STATA 15 Output

Table 6: Impact of Trade Facilitation on Economic Growth among Middle-Income Countries

| | With the Effects of Logistic Performance Indices | | | | With the Effect of Overall Logistic Performance Index | | | |
|---|---|---|---|---|---|---|---|---|
| | (1) | (2) | (3) | (4) | (5) | (6) | (7) | (8) |
| VARIABLES | One-step Difference GMM | Two-step Difference GMM | One-step System GMM | Two-step System GMM | One-step Difference GMM | Two-step Difference GMM | One-step System GMM | Two-step System GMM |
| | lnGNI | lnGNI | lnGNI | lnGNI | lnGNI | lnGNI | lnGNI | lnGNI |
| L.lnGNI | 0.640*** | 0.530* | 0.644*** | 0.655*** | 0.686*** | 0.634** | 0.730*** | 0.853*** |
| | (0.234) | (0.312) | (0.187) | (0.246) | (0.197) | (0.279) | (0.157) | (0.175) |
| ATCE | 0.000244 | 0.000360 | -0.00399 | -0.00409 | 0.000165 | 0.000137 | -0.00310 | -0.00144 |
| | (0.00144) | (0.00178) | (0.00434) | (0.00523) | (0.00133) | (0.00158) | (0.00321) | (0.00350) |
| lnPCT | -0.00875 | -0.00836 | -0.0356* | -0.0371 | -0.00835 | -0.00670 | -0.0254 | -0.0115 |
| | (0.0115) | (0.00774) | (0.0209) | (0.0321) | (0.0105) | (0.00646) | (0.0164) | (0.0183) |
| AFT | -4.42e-06 | -1.88e-06 | 7.93e-07 | 1.70e-06 | -4.84e-06 | -3.77e-06 | 1.87e-06 | 1.18e-06 |
| | (4.31e-06) | (5.71e-06) | (5.53e-06) | (5.41e-06) | (3.92e-06) | (5.01e-06) | (4.43e-06) | (4.01e-06) |
| QPI | -0.00187 | -0.00840 | 0.0251 | 0.0287 | 0.00207 | -0.000437 | 0.0216 | 0.0183 |
| | (0.0153) | (0.0210) | (0.0206) | (0.0252) | (0.0127) | (0.0171) | (0.0171) | (0.0228) |
| LPIAC | 0.0330 | 0.0458 | 0.0181 | 0.00980 | | | | |
| | (0.0333) | (0.0373) | (0.0429) | (0.0445) | | | | |
| LPICQ | 0.0156 | 0.0125 | 0.00441 | 0.0214 | | | | |
| | (0.0311) | (0.0430) | (0.0543) | (0.0621) | | | | |
| LPIEA | -0.00487 | -0.0170 | -0.0564 | -0.0704 | | | | |
| | (0.0177) | (0.0235) | (0.0571) | (0.0672) | | | | |
| LPIEC | 0.0387* | 0.0464** | -0.0455 | -0.0255 | | | | |
| | (0.0201) | (0.0232) | (0.0611) | (0.0575) | | | | |
| LPIFS | -0.0378 | -0.0350 | 0.0124 | -0.00824 | | | | |
| | (0.0341) | (0.0333) | (0.0375) | (0.0439) | | | | |
| LPIQTT | -0.0625* | -0.0792** | 0.144 | 0.149 | | | | |
| | (0.0370) | (0.0384) | (0.100) | (0.126) | | | | |
| TRF | 7.03e-06 | 0.000736 | -0.0106** | -0.00734 | -0.000537 | 0.000237 | -0.00802* | -0.00352 |
| | (0.00146) | (0.00175) | (0.00524) | (0.00596) | (0.00128) | (0.00165) | (0.00431) | (0.00485) |
| lnEXPG | 0.0355 | 0.0413 | 0.0434 | 0.0472 | 0.0364 | 0.0422 | 0.0335 | 0.0273 |
| | (0.0299) | (0.0408) | (0.0317) | (0.0410) | (0.0289) | (0.0385) | (0.0255) | (0.0326) |
| lnIMPG | -0.00453 | -0.00127 | 0.0107 | -0.00255 | -0.00963 | -0.00940 | 0.00779 | -0.0147 |
| | (0.0267) | (0.0365) | (0.0413) | (0.0447) | (0.0272) | (0.0357) | (0.0322) | (0.0337) |
| lnFDI | 0.0158 | 0.0186 | 0.0340 | 0.0292 | 0.0145 | 0.0180* | 0.0196 | 0.00548 |
| | (0.0106) | (0.0113) | (0.0359) | (0.0390) | (0.00953) | (0.0100) | (0.0264) | (0.0242) |
| lnGFCF | 0.0425* | 0.0556 | -0.000799 | 0.00209 | 0.0367* | 0.0455 | -0.000424 | 0.00333 |
| | (0.0242) | (0.0369) | (0.0138) | (0.0181) | (0.0205) | (0.0316) | (0.0103) | (0.0153) |
| yr1 | 0 | 0 | 0 | 0 | 0 | 0 | 0 | 0 |
| | (0) | (0) | (0) | (0) | (0) | (0) | (0) | (0) |
| yr2 | -0.0161 | -0.0497 | -0.0502 | -0.0313 | -0.00304 | -0.0173 | 1.151 | 0.0457 |
| | (0.0820) | (0.103) | (0.0400) | (0.0929) | (0.0703) | (0.0923) | (0.904) | (0.0593) |
| yr3 | 0.00734 | -0.0331 | -0.0356 | -0.0298 | 0.0197 | -0.00355 | 1.171 | 0.0528 |
| | (0.0655) | (0.0899) | (0.0352) | (0.0904) | (0.0528) | (0.0786) | (0.911) | (0.0516) |
| yr4 | 0.0161 | -0.0149 | -0.0227 | -0.0158 | 0.0285 | 0.0112 | 1.176 | 0.0545 |
| | (0.0587) | (0.0767) | (0.0263) | (0.0797) | (0.0468) | (0.0655) | (0.914) | (0.0459) |
| yr5 | 0.0168 | -0.00721 | -0.0195 | -0.00884 | 0.0293 | 0.0169 | 1.173 | 0.0503 |
| | (0.0513) | (0.0657) | (0.0169) | (0.0686) | (0.0399) | (0.0546) | (0.919) | (0.0392) |
| yr6 | 0.0107 | -0.00541 | -0.0221*** | -0.0130 | 0.0194 | 0.0114 | 1.166 | 0.0342 |
| | (0.0385) | (0.0517) | (0.00748) | (0.0562) | (0.0304) | (0.0425) | (0.925) | (0.0289) |
| yr7 | 0.0334 | 0.0202 | 0 | 0.0114 | 0.0397 | 0.0332 | 1.188 | 0.0569** |
| | (0.0353) | (0.0478) | (0) | (0.0516) | (0.0285) | (0.0401) | (0.926) | (0.0256) |
| yr8 | 0.0604* | 0.0426 | 0.0272*** | 0.0359 | 0.0668*** | 0.0554 | 1.214 | 0.0758*** |
| | (0.0310) | (0.0408) | (0.00703) | (0.0451) | (0.0248) | (0.0354) | (0.928) | (0.0233) |
| yr9 | 0.0699*** | 0.0566* | 0.0351*** | 0.0457 | 0.0755*** | 0.0675*** | 1.219 | 0.0793*** |
| | (0.0224) | (0.0306) | (0.0108) | (0.0389) | (0.0174) | (0.0250) | (0.932) | (0.0189) |
| yr10 | 0.0721*** | 0.0640*** | 0.0442* | 0.0602*** | 0.0748*** | 0.0694*** | 1.219 | 0.0758*** |
| | (0.0131) | (0.0169) | (0.0228) | (0.0215) | (0.0113) | (0.0142) | (0.937) | (0.0120) |
| yr11 | 0 | 0 | -0.0202 | 0 | 0 | 0 | 1.146 | 0 |
| | (0) | (0) | (0.0310) | (0) | (0) | (0) | (0.938) | (0) |
| LPI | | | | | -0.0210 | -0.0299 | 0.0638 | 0.0345 |
| | | | | | (0.0188) | (0.0185) | (0.0481) | (0.0551) |
| Constant | | | 1.430 | 1.617 | | | 0 | 0.839 |
| | | | (1.155) | (1.404) | | | (0) | (1.052) |
| Observations | 846 | 846 | 940 | 940 | 846 | 846 | 940 | 940 |
| Number of country | 94 | 94 | 94 | 94 | 94 | 94 | 94 | 94 |
| No of Instruments | 33 | 33 | 35 | 35 | 28 | 28 | 30 | 30 |

Source: Extractions from STATA 15 Output

Table 7: Impact of Trade Facilitation on Trade Flows and Economic Growth among Lower-middle-income countries

| | lnEXPG with & without individual logistic performance | | lnIMPG with & without individual logistic performance | | lnGNI with & without individual logistic performance | |
|---|---|---|---|---|---|---|
| | (1) | (2) | (3) | (4) | (5) | (6) |
| VARIABLES | lnEXPG | lnEXPG | lnIMPG | lnIMPG | lnGNI | lnGNI |
| L.lnEXPG | 0.602* | 0.561 | | | | |
| | (0.328) | (0.379) | | | | |
| ATCE | 0.00294 | -0.000402 | -0.0115 | -0.0145 | -0.00189 | -0.00201 |
| | (0.0126) | (0.0134) | (0.0114) | (0.0126) | (0.00299) | (0.00332) |
| lnPCT | 0.194 | 0.215 | 0.150* | 0.154* | -0.0111 | -0.0112 |
| | (0.140) | (0.160) | (0.0882) | (0.0913) | (0.0162) | (0.0166) |
| AFT | -0.000364 | -0.000440 | -0.000239 | -0.000249 | 3.75e-06 | 4.40e-06 |
| | (0.000422) | (0.000492) | (0.000207) | (0.000208) | (1.40e-05) | (1.54e-05) |
| QPI | -0.0406 | -0.0478 | -0.0487 | -0.0537 | -0.0169 | -0.0179 |
| | (0.0625) | (0.0698) | (0.0467) | (0.0469) | (0.0228) | (0.0232) |
| LPIAC | -0.118 | | -0.0198 | | 0.0366 | |
| | (0.205) | | (0.165) | | (0.0424) | |
| LPICQ | 0.150 | | 0.100 | | 0.0375 | |
| | (0.234) | | (0.183) | | (0.0718) | |
| LPIEA | 0.0699 | | 0.220 | | -0.0244 | |
| | (0.266) | | (0.219) | | (0.0328) | |
| LPIEC | -0.290 | | -0.339 | | 0.0295 | |
| | (0.277) | | (0.248) | | (0.0367) | |
| LPIFS | 0.279 | | 0.188 | | -0.0135 | |
| | (0.257) | | (0.128) | | (0.0311) | |
| LPIQTT | 0.0193 | | 0.202 | | -0.0275 | |
| | (0.192) | | (0.131) | | (0.0479) | |
| TRF | -0.0128 | -0.0146 | -0.0178 | -0.0184 | -0.00412 | -0.00446 |
| | (0.0190) | (0.0234) | (0.0165) | (0.0185) | (0.00385) | (0.00413) |
| lnFDI | -11.29* | -12.17* | -9.639*** | -9.706*** | 0.339 | 0.349 |
| | (5.899) | (6.570) | (2.977) | (2.973) | (0.382) | (0.432) |
| lnGFCF | 0.168 | 0.201 | 0.147 | 0.158 | 0.0146 | 0.0166 |
| | (0.229) | (0.272) | (0.165) | (0.177) | (0.0133) | (0.0147) |
| yr1 | 0 | 0 | 0 | 0 | 0 | 0 |
| | (0) | (0) | (0) | (0) | (0) | (0) |
| yr2 | 0.0688 | 321.4* | 0.227** | 0.267*** | -7.471 | -7.617 |
| | (0.0875) | (172.6) | (0.0894) | (0.0734) | (9.618) | (10.83) |
| yr3 | 0 | 321.3* | 0.197*** | 0.207*** | -7.461 | -7.602 |
| | (0) | (172.6) | (0.0550) | (0.0479) | (9.612) | (10.82) |
| yr4 | 0.0178 | 321.3* | 0.203*** | 0.208*** | -7.453 | -7.592 |
| | (0.0836) | (172.6) | (0.0473) | (0.0456) | (9.612) | (10.82) |
| yr5 | 0.00437 | 321.2* | 0.153*** | 0.151*** | -7.456 | -7.593 |
| | (0.0670) | (172.6) | (0.0313) | (0.0274) | (9.610) | (10.82) |
| yr6 | -0.223*** | 321.0* | 0 | 0 | -7.463 | -7.601 |
| | (0.0787) | (172.6) | (0) | (0) | (9.604) | (10.81) |
| yr7 | -0.154 | 321.1* | 0.0149 | 0.0164 | -7.444 | -7.584 |
| | (0.0925) | (172.5) | (0.0396) | (0.0440) | (9.602) | (10.81) |
| yr8 | -0.0178 | 321.2* | 0.151*** | 0.147** | -7.426 | -7.564 |
| | (0.0948) | (172.5) | (0.0531) | (0.0567) | (9.602) | (10.81) |
| yr9 | -0.00158 | 321.2* | 0.183*** | 0.173*** | -7.416 | -7.552 |
| | (0.0783) | (172.6) | (0.0429) | (0.0422) | (9.600) | (10.81) |
| yr10 | -0.106 | 321.1* | 0.0706 | 0.0768 | -7.421 | -7.557 |
| | (0.107) | (172.5) | (0.0536) | (0.0476) | (9.596) | (10.80) |
| yr11 | -0.0326 | 321.2* | 0.0915 | 0.113 | -7.475 | -7.614 |
| | (0.150) | (172.6) | (0.112) | (0.104) | (9.593) | (10.80) |
| LPI | | 0.140 | | 0.382* | | 0.0409 |
| | | (0.237) | | (0.207) | | (0.0545) |
| L.lnIMPG | | | 0.554* | 0.551* | | |
| | | | (0.296) | (0.304) | | |
| L.lnGNI | | | | | 0.733*** | 0.714*** |
| | | | | | (0.186) | (0.204) |
| lnEXPG | | | | | 0.0133 | 0.0163 |
| | | | | | (0.0170) | (0.0181) |
| lnIMPG | | | | | 0.0179 | 0.0141 |
| | | | | | (0.0279) | (0.0279) |
| Constant | 298.1* | 0 | 256.6*** | 258.2*** | 0 | 0 |
| | (155.0) | (0) | (79.39) | (79.20) | (0) | (0) |
| | | | | | | |
| Observations | 480 | 480 | 480 | 480 | 480 | 480 |
| Number of country | 48 | 48 | 48 | 48 | 48 | 48 |
| No of Instruments | 33 | 28 | 33 | 28 | 35 | 30 |

Source: Extractions from STATA 15 Output

Table 8: Impact of Trade Facilitation on Trade Flows and Economic Growth among Upper-middle-income countries

| | lnEXPG with & without individual logistic performance | | lnIMPG with & without individual logistic performance | | lnGNI with & without individual logistic performance | |
|---|---|---|---|---|---|---|
| | (1) | (2) | (3) | (4) | (5) | (6) |
| VARIABLES | lnEXPG | lnEXPG | lnIMPG | lnIMPG | lnGNI | lnGNI |
| L.lnEXPG | 0.828*** | 0.856*** | | | | |
| | (0.0983) | (0.0871) | | | | |
| ATCE | -0.00657 | -0.00638 | -0.00841 | -0.00630 | -0.00265 | -0.00225 |
| | (0.00489) | (0.00429) | (0.00817) | (0.00674) | (0.00372) | (0.00276) |
| lnPCT | 0.0230 | 0.0222 | 0.0102 | 0.00956 | -0.00480 | -0.00420 |
| | (0.0173) | (0.0161) | (0.0176) | (0.0171) | (0.0103) | (0.00774) |
| AFT | -7.29e-06 | -5.16e-06 | -1.20e-09 | 7.68e-07 | -1.07e-06 | -1.26e-06 |
| | (5.80e-06) | (5.43e-06) | (3.83e-06) | (3.17e-06) | (5.09e-06) | (5.06e-06) |
| QPI | 0.00877 | 0.00555 | 0.0181 | 0.0125 | 0.0168 | 0.0138 |
| | (0.0260) | (0.0276) | (0.0194) | (0.0192) | (0.0201) | (0.0144) |
| LPIAC | -0.0322 | | 0.105 | | 0.0424 | |
| | (0.114) | | (0.130) | | (0.109) | |
| LPICQ | 0.103 | | 0.0336 | | 0.0353 | |
| | (0.119) | | (0.105) | | (0.0311) | |
| LPIEA | -0.0193 | | -0.0320 | | 0.00295 | |
| | (0.103) | | (0.0619) | | (0.0329) | |
| LPIEC | -0.299* | | -0.220 | | -0.0299 | |
| | (0.166) | | (0.175) | | (0.0949) | |
| LPIFS | 0.0959* | | 0.0422 | | -0.00207 | |
| | (0.0519) | | (0.0414) | | (0.0270) | |
| LPIQTT | 0.264** | | 0.211** | | -0.0140 | |
| | (0.108) | | (0.100) | | (0.0425) | |
| TRF | -0.00546 | -0.00430 | -0.00552 | -0.00367 | -0.00191 | -0.00195 |
| | (0.00445) | (0.00399) | (0.00496) | (0.00457) | (0.00171) | (0.00147) |
| lnFDI | -0.0118 | -0.0132 | -0.0399 | -0.0385 | 0.0141 | 0.0110 |
| | (0.0249) | (0.0236) | (0.0240) | (0.0261) | (0.0464) | (0.0408) |
| lnGFCF | 0.124 | 0.106 | 0.0569 | 0.0315 | 0.00952 | 0.00958 |
| | (0.0919) | (0.0824) | (0.101) | (0.0903) | (0.0244) | (0.0240) |
| yr1 | 0 | 0 | 0 | 0 | 0 | 0 |
| | (0) | (0) | (0) | (0) | (0) | (0) |
| yr2 | 0.963 | 0.754 | 0.0652 | 1.801 | 0.0447 | 0.866 |
| | (0.859) | (0.752) | (0.152) | (1.207) | (0.134) | (1.083) |
| yr3 | 0.820 | 0.613 | -0.0566 | 1.679 | 0.0742 | 0.893 |
| | (0.848) | (0.747) | (0.127) | (1.226) | (0.0931) | (1.122) |
| yr4 | 0.810 | 0.589 | -0.0559 | 1.666 | 0.0692 | 0.885 |
| | (0.863) | (0.752) | (0.123) | (1.224) | (0.0845) | (1.121) |
| yr5 | 0.768 | 0.534 | -0.0892 | 1.618 | 0.0588 | 0.872 |
| | (0.874) | (0.755) | (0.113) | (1.228) | (0.0681) | (1.130) |
| yr6 | 0.622 | 0.386 | -0.220** | 1.483 | 0.0375 | 0.849 |
| | (0.871) | (0.755) | (0.107) | (1.239) | (0.0507) | (1.146) |
| yr7 | 0.773 | 0.540 | -0.139 | 1.569 | 0.0649 | 0.875 |
| | (0.874) | (0.758) | (0.123) | (1.219) | (0.0478) | (1.147) |
| yr8 | 0.908 | 0.681 | 0.00558 | 1.723 | 0.0986** | 0.909 |
| | (0.861) | (0.751) | (0.140) | (1.203) | (0.0407) | (1.156) |
| yr9 | 0.895 | 0.667 | 0.0191 | 1.736 | 0.0984*** | 0.908 |
| | (0.860) | (0.748) | (0.130) | (1.215) | (0.0268) | (1.170) |
| yr10 | 0.827 | 0.601 | -0.0361 | 1.678 | 0.0968*** | 0.907 |
| | (0.857) | (0.752) | (0.112) | (1.237) | (0.0181) | (1.184) |
| yr11 | 0.935 | 0.712 | 0 | 1.712 | 0 | 0.811 |
| | (0.931) | (0.820) | (0) | (1.328) | (0) | (1.194) |
| LPI | | 0.147 | | 0.167* | | 0.0339 |
| | | (0.101) | | (0.0977) | | (0.0564) |
| L.lnIMPG | | | 0.873*** | 0.918*** | | |
| | | | (0.156) | (0.140) | | |
| L.lnGNI | | | | | 0.820** | 0.851*** |
| | | | | | (0.338) | (0.287) |
| Constant | 0 | 0 | 2.257 | 0 | 1.017 | 0 |
| | (0) | (0) | (1.501) | (0) | (1.565) | (0) |
| Observations | 460 | 460 | 460 | 460 | 460 | 460 |
| Number of country | 46 | 46 | 46 | 46 | 46 | 46 |
| No of Instruments | 33 | 28 | 33 | 28 | 35 | 30 |

Source: Extractions from STATA 15 Output